\documentclass[11pt]{article}
\usepackage{bm}
\usepackage[normalem]{ulem}
\usepackage{amsmath,graphics,epsfig,float,braket,multirow,url}
\usepackage{caption}
\usepackage{subcaption}
\usepackage[toc,page]{appendix}
\usepackage{comment}
\usepackage{authblk}
\usepackage{hyperref}
\usepackage[capitalize]{cleveref}
\usepackage[normalem]{ulem}
\usepackage{diffcoeff}
\usepackage{xcolor, soul}
\usepackage{multirow}
\usepackage{booktabs}

\newcommand{\cosec}{\operatorname{cosec}}

\newcommand{\s}[1]{{\textsf{\textbf{#1}}}}

\begin{document}

\title{\s{Design and modelling of compliant mechanisms with invertible Poisson's ratio effect for growing biological cells}}
\author{
\textsf{Manu Sebastian$^{\dagger}$, Sreenath Balakrishnan$^{\ddagger}$, and Safvan Palathingal$^{\dagger}$}
}

\date{
{\it $^{\dagger}$Department of Mechanical and Aerospace Engineering,\\
Indian Institute of Technology Hyderabad, Telangana, India}\\[1ex]
{\it $^{\ddagger}$School of Mechanical Sciences,\\
Indian Institute of Technology Goa, Goa, India}\\[2ex]
August 20, 2023
}

 \maketitle
\hrule\vskip 6pt

\begin{abstract}
 The behaviour of biological cells depends on the mechanical properties, such as Elastic Modulus and Poisson’s ratio, of the substrate they adhere to. Tunable materials such as polyacrylamide gels and hydrogels were previously used as substrates to understand this dependence. However, these substrates do not facilitate changing their elastic properties in situ while cells are growing on them. This work presents an alternate approach that enables this---substrates based on tunable compliant micro mechanisms.
In particular, the mechanism proposed here has an invertible Poisson’s ratio effect. In the first configuration, the effect is positive, and in the second, it is negative, with any desired magnitude. We achieve this by changing the stiffness between two internal points of a mechanism with the shape of a re-entrant structure. An increase in stiffness causes the direction of deformation along the lateral axis to reverse for a given reference load along the horizontal axis.
We derive analytical expressions that relate the geometric parameters to the ratio of input and output displacements for both mechanism configurations. The analytical modelling is verified with finite element analysis and experiments on mesoscale design prototypes of both configurations.
    
\end{abstract}

\vskip 6pt
\hrule
\vskip 6pt
\section{Introduction}

\begin{figure*}[ht]
    \centering\includegraphics[width=0.7\linewidth]{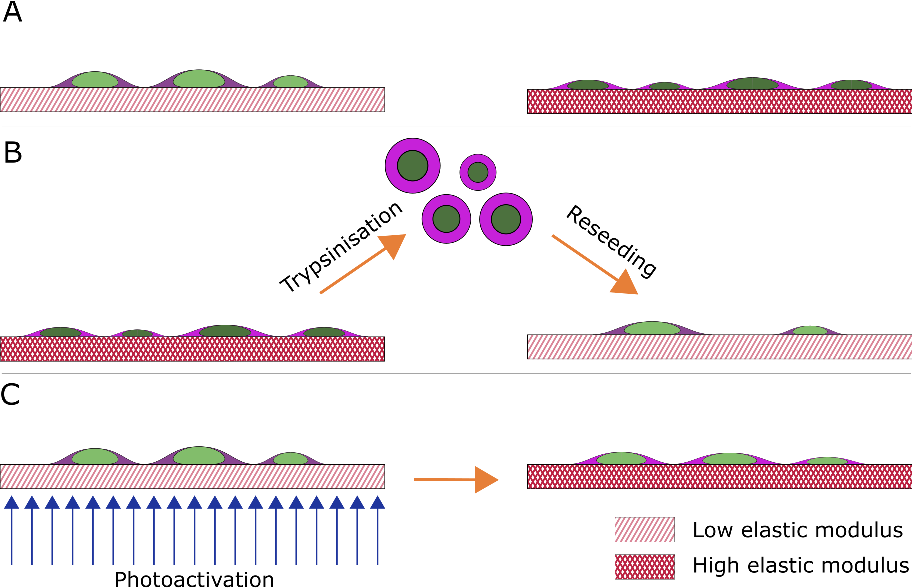}
    \caption{Current methods for studying the cell response to alterations in mechanical properties of the substrate. (A) Cells are grown on substrates with different elastic modulus, and the differences in cell properties between them are noted. (B) Cells are transferred from one substrate to another by trypsinizing (clipping the protein linkages between the cell and its substrate) from one and reseeding into the other. (C) The Elastic modulus of light-sensitive substrates is altered by photoactivation while cells grow on them.}
    \label{fig:intro}
\end{figure*}

Biological cells are known to respond to the mechanical properties of their environment. When stem cells were grown on substrates of different elastic modulus (Fig.~\ref{fig:intro}), 
\begin{figure*}[!htbp]
\centering\includegraphics[width=0.9\linewidth]{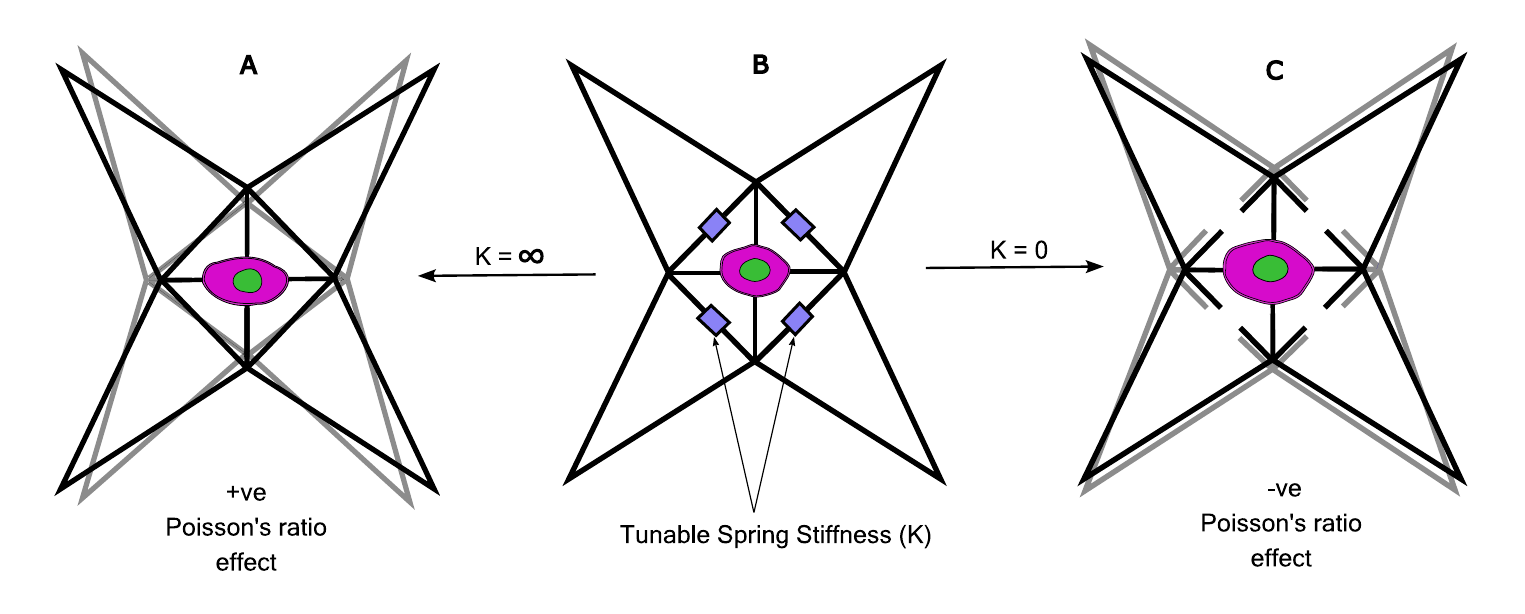}
\caption{Compliant mechanism with tunable Poisson's ratio}
\label{fig:tunable_mechanism_schematic}
\end{figure*}
the ones on the softer substrate converted into brain-like cells, whereas the ones on the stiffer substrate converted into bone-like cells, conforming to the elastic properties of these tissues; the brain is soft and bone is stiff \cite{Engler2006}. This seminal observation led to broad interest in unravelling the relationship between cell function and the elastic properties of its substrate. First studies were conducted by creating soft substrates (elastic modulus in the range of 1 - 100 kPa) using polyacrylamide gels \cite{Engler2006}, hydrogels \cite{DRURY20034337} or PDMS pillars \cite{Saez2007}, growing cells on them and subsequently comparing the difference in cell behaviour on these substrates (Fig.~\ref{fig:intro}A). Even though these studies are very insightful, they do not capture the fact that the extracellular matrix to which the cells adhere is continuously remodelled due to various factors such as ageing, inflammation, fibrosis and diseases such as cancer. Several new techniques were invented to study the response of cells to this evolving nature of the extracellular matrix. Cells were detached (by trypsinization) from substrates with high elastic modulus and transferred to substrates with low elastic modulus (Fig.~\ref{fig:intro}B) \cite{Yang2014}. Such techniques caused difficulties in interpreting the results due to the detachment and re-attachment of cells during substrate switching. Photoactivatable hydrogels were developed to overcome these complications; the elastic modulus of these gels could be modified by irradiation while cells are growing on them (Fig.~\ref{fig:intro}C) \cite{Kloxin2009}. {In comparison to the studies mentioned above on the effect of elastic modulus on cell function, fewer studies have examined the effect of Poisson's ratio on cell function }\cite{Yan2017,Song2018,Zhang2013}. {This is probably because---(i) cells respond to mechanical properties of substrate only at low values of elastic modulus (1 - 100 kPa) and (ii) material design techniques typically employed for creating substrates with low elastic modulus, for example, gels and pillars, are not amenable to manipulation of Poisson's ratio. In }\cite{Yan2017,Song2018}{, the authors have used poly-urethane scaffolds to show that negative Poisson's ratio enhances vascular and neuronal differentiation of pluripotent stem cells. In }\cite{Zhang2013}{, the authors have fabricated a re-entrant structure using two-photon polymerization and showed that negative Poisson's ratio induces abnormalities in cell division.  However, to the best of our knowledge, there are no known methods for altering the Poisson's ratio of the substrate while the cells are growing on them. Other limitations of the current materials-based techniques for substrate modification are---(i) multiple switching back and forth between elastic moduli, i.e., high-low-high-low etc., is difficult, and (ii) the range of elastic properties between which switching is possible is limited.}

To resolve these limitations, we have developed an alternate framework for modifying the elastic properties of the substrate while the cells are growing on them---compliant micromechanisms with tunable stiffness elements. {Cells attach to multiple points in its microenvironment through protein complexes known as focal adhesions} \cite{Geiger2009}. {By pulling and pushing at these finite attachment points, the cells probe the elastic properties of its microenvironment} \cite{Plotnikov2012}. Therefore, the cells respond to the \textit{force-displacement relationship} between the focal adhesions and not directly to the elastic properties of its substrate, such as Young's modulus and Poisson's ratio. Hence, we can mimic an equivalent substrate elastic property by designing a mechanism between these attachment points, which gives a required force-displacement relationship. Here, we design a cell substrate which can be switched between positive and negative Poisson's ratio by integrating tunable stiffness elements into a compliant mechanism. Previously, we had designed a single-cell, biaxial stretcher based on re-entrant structures \cite{fartyal2021} (Fig.~\ref{fig:tunable_mechanism_schematic}C). On this mechanism, if the cells push outward in the horizontal direction, the vertical direction also moves outward, thereby giving the cell a sense of adhering to a negative Poisson's ratio substrate. Now if the re-entrant points are connected (Fig.~\ref{fig:tunable_mechanism_schematic}A), the cell feels a sense of adhering to a substrate with a positive Poisson's ratio. By using tunable stiffness elements (e.g., bistable elements) between the re-entrant points (Fig.~\ref{fig:tunable_mechanism_schematic}B), we can alternate between these configurations.      

\begin{figure}[!htbp]
    \centering
    \includegraphics[width=0.3\linewidth]{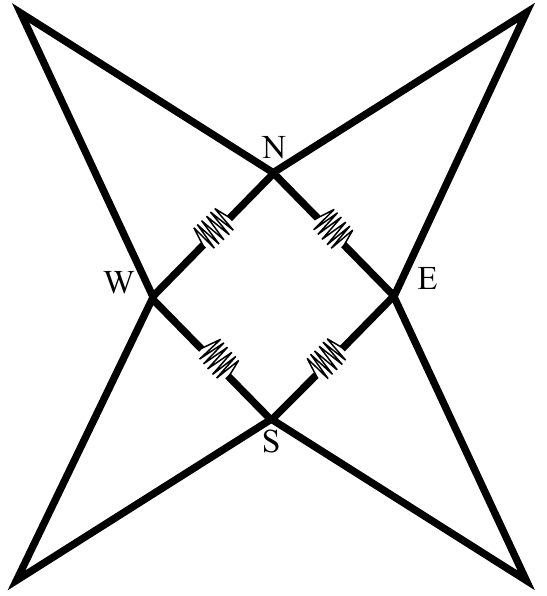}
    \caption{Frame with tunable springs}
    \label{fig:frame_with_springs}
\end{figure}

In Sec. \ref{sec:modelling}, we model the mechanism with invertible Poisson's ratio  (see Fig. \ref{fig:frame_with_springs}). 
and derive a relationship between the deformation of the two re-entrant points (points N and E) and spring stiffness. We do not lose generality by selecting only these two points, as the mechanism is symmetric.  In this work, we consider the two extremes of stiffness  (zero and infinity) and obtain expressions for the stretch ratio between points N and E. By using these stretch ratios, we can design mechanisms that can switch between two given positive and negative Poisson's ratio. It involves the design of these two configurations and that of an element with tunable stiffness that switches between zero and infinity.  We illustrate the design of two configurations, one with no connection between points N and E and another with a rigid connection between points N and E, for a given positive and negative stretch ratios with the help of two case studies in Sec. \ref{sec:validation}. We validate these two designs with prototypes and finite element analysis (FEA). The design of the tunable element that switches the mechanism between the two configurations is not in this work's scope. However, we discuss a possible design for the tunable element by using a bistable element in Sec. \ref{sec:summary}, which is a future direction of this work.

\section{Analytical modelling}
\label{sec:modelling}
As  described in the previous section, a linear spring of variable stiffness $K$, as shown in Fig. \ref{fig:frame_with_springs}, is the key mechanical element responsible for inverting the Poisson's ratio effect. The case of disengaged configuration is modelled by taking $K=0$ and engaged configuration with a finite and large value for $K$.

\begin{figure}[!htbp]
    \centering
    \includegraphics[width=0.4\linewidth]{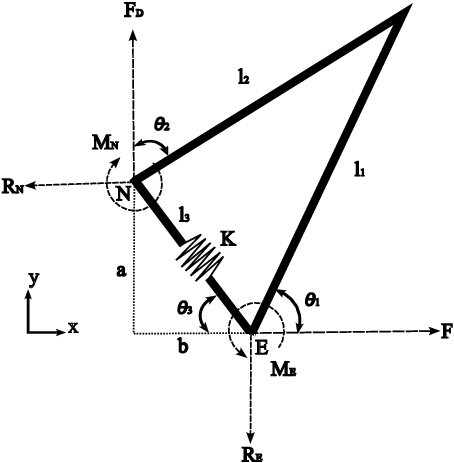}
    \caption{Quarter of the mechanism with Force, $F$, at point E and a dummy Load, $F_D$, at point N.}
    \label{fig:quarter_frame}
\end{figure}

 In order to design mechanisms for a given Poisson's ratio effect, we intend to find the relationship between the displacements at points N and E.   We use the symmetry in the mechanism for simplifying the analysis. The mechanism is analysed by considering quarter of it as depicted in Fig. \ref{fig:quarter_frame}.

\subsection{Force and moment balance}
A point load $F$ is applied at the point E, and a dummy load  $F_D$ at the point N, as shown in  Fig. \ref{fig:quarter_frame}. Reactions $R_N$, $R_E$, $M_E$, and $M_N$ can be obtained from the force and moment balance as
\begin{align}
    R_N &= F,\\
    R_E &= F_D,\\
    M_{N} &= M_E + Fa - F_Db,
\end{align}
where a and b are the y-coordinate of N and x-coordinate of E, respectively.

\subsection{Total strain energy}
\begin{figure}[!htbp]
    \centering
    \includegraphics[width=0.4\linewidth]{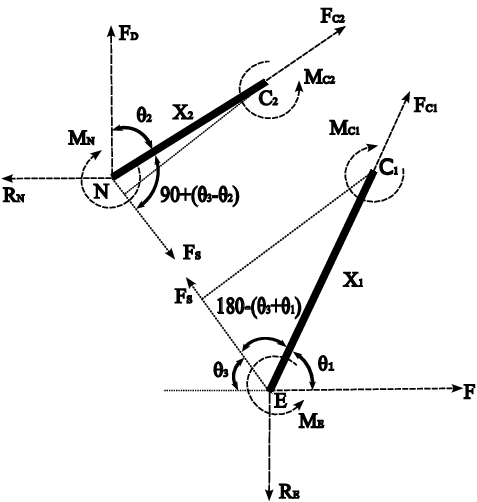}
    \caption{Cut sections of frame}    
    \label{fig:internal_forces}
\end{figure}

The reaction moment at a  cross-section  can be obtained in terms of the geometrical and material properties of the mechanism. We consider the free body diagrams given in Fig. \ref{fig:internal_forces}.  
The moment, $M_{C_1}$, in the member $\mathrm{EC_1}$ at the Point $\mathrm{C_1}$ is,
\begin{equation}
    M_{C_1} = M_E +F_DX_1cos\theta_1 + FX_1sin\theta_1 - F_S X_1 sin(\theta_1+\theta_3),
    \label{Eq:MC1}
\end{equation}
where $X_1$ is the distance between  points E and $\mathrm{C_1}$, and  $F_S$ is the reaction force in the spring. Similarly,  moment $M_{C_2}$, in the member $\mathrm{NC_2}$ is,
\begin{multline}    \label{Eq:MC2}
    M_{C_2} = M_E +Fa - F_Db + FX_2cos(\theta_2) +F_DX_2sin(\theta_2) - F_SX_2cos(\theta_3-\theta_2),
\end{multline}
where $X_2$ is the distance between  points N and $\mathrm{C_2}$.

By using the Euler-Bernoulli beam theory, the bending strain energy of the mechanism is 
\begin{equation}
    SE_B =   \int\limits_0^{l_1} \frac{M_{C_1}^2}{2YI} \ dX_1 + \int\limits_0^{l_2} \frac{M_{C_2}^2}{2YI} \ dX_2,
\end{equation}
where $l_1$ and $l_2$ are the lengths of the bottom and top members as shown in Fig. \ref{fig:quarter_frame}, $Y$ is the modulus of elasticity, and $I$ is the second  moment of the area.
The strain energy in the  spring, $SE_S$, is,
\begin{equation}
    SE_S = \frac{F_S^2}{2K}
    \label{Eq:SEs}
\end{equation}

The total strain energy, $SE$, is the sum of $SE_B$ and $SE_S$, which on integrating and simplifying, we have,
\begin{multline} \label{Eq:SE}
SE = \frac{(a F-b W+F l_2 \cos (\theta_2)-l_2 S \cos (\theta_3-\theta_2)}{3 (F \cos (\theta_2)-S \cos (\theta_3-\theta_2)+W \sin (\theta_2))}
+ \frac{l_2 W \sin (\theta_2)+M_E)^3-(a F-b W+M_E)^3}{3 (F \cos (\theta_2)-S \cos (\theta_3-\theta_2)+W \sin (\theta_2))}\\
+\frac{(F l_1 \sin (\theta_1)-l_1 s \sin (\theta_1+\theta_3)+l_1 W \cos (\theta_1)+M_E)^3-M_E^3}{3 (F \sin (\theta_1)-S \sin (\theta_1+\theta_3)+W \cos (\theta_1))}
+\frac{F_S^2}{2K}
\end{multline}

\subsection{Displacement at points E and N}
Next, we express $M_E$ in Eq. \ref{Eq:SE} in terms of the applied load and dummy loads so that the final expression of the strain energy becomes amenable for application of Castigliano's Theorem. 

Due to the symmetry of the mechanism, the rotation of the cross section about the Point E should be zero. Therefore, 
\begin{equation}
\frac{dSE}{dM_E} = 0,
\end{equation}
which implies that,
\begin{multline} \label{Eq:ME}
 M_E = \frac{F( -2 a l_2-l_1^2\sin (\theta_1)-l_2^2\cos(\theta_2)}{2(l_1+l_2)}
 +\frac{F_D \left(2 b l_2-l_1^2 \cos (\theta_1)-l_2^2 \sin (\theta_2)\right)}{2(l_1+l_2)}\\
 +\frac{F_S(-l_1^2  \sin (\theta_1+\theta_3)-l_2^2 \cos (\theta_3-\theta_2))}{2 (l_1+l_2)}.
\end{multline}
For brevity, we define the factor multiplying $F$, $F_D$, $F_S$ in Eq. \ref{Eq:ME} as $\alpha$, $\beta$, and $\gamma$, respectively, so that  Eq. \ref{Eq:ME} becomes,
\begin{equation} \label{Eq:ME_short}
 M_E = \alpha F + \beta F_D + \gamma F_S.
\end{equation}
Substituting this expression of $M_E$ in Eq. \ref{Eq:SE}, differentiating $SE$ with respect to $F$ and equating the dummy load to zero, we get the displacement $\delta_X$ at point E, as
\begin{equation}\label{Eq:delta_x}
\begin{split}
     \delta_X &= \diffp{SE}F[W=0]\\
       &=\frac {3 (a + \alpha + l_2\cos (\theta_2)) (a F + \alpha F + 
      Fl_2\cos (\theta_2) + \gamma F_S}{6YI (F\cos (\theta_2) - 
    F_S\cos (\theta_3 - \theta_2))}\\
    &-\frac{ l_2 F_S\cos (\theta_3 - \theta_2))^2 - 
  3 (a + \alpha) (a F + \alpha F + \gamma F_S)^2} {6YI (F\cos (\theta_2) - 
    F_S\cos (\theta_3 - \theta_2))}\\
    &- \frac {\cos (\theta_2)(a F + \alpha F + 
       Fl_2\cos (\theta_2) + \gamma F_S - l_2 F_S\cos (\theta_3 - \theta_2))^3}{6YI (F\cos (\theta_2) - 
     F_S\cos (\theta_3 - \theta_2))^2} \\
      &-\frac{\cos (\theta_2)(a F + \alpha F + \gamma F_S)^3} {6YI (F\cos (\theta_2) - 
     F_S\cos (\theta_3 - \theta_2))^2} \\
     &+ \frac {3 (\alpha + l_1\sin (\theta_1)) (\alpha F + 
      Fl_1\sin (\theta_1) + \gamma F_S - l_1 F_S\sin (\theta_1 + \theta_3))^2}{6YI (F\sin (\theta_1) - 
    F_S\sin (\theta_1 + \theta_3))} \\ 
  &-\frac{3\alpha (\alpha F + \gamma F_S)^2} {6YI (F\sin (\theta_1) - 
    F_S\sin (\theta_1 + \theta_3))} \\
    &- \frac {\sin (\theta_1)((\alpha F + 
       Fl_1\sin (\theta_1) + \gamma F_S - l_1 F_S\sin (\theta_1 + \theta_3))^3}{6YI (F\sin (\theta_1) - F_S\sin (\theta_1+ \theta_3))^2}\\
       &-\frac{\sin (\theta_1)(\alpha F + \gamma F_S)^3 } {6YI (F\sin (\theta_1) - F_S\sin (\theta_1+ \theta_3))^2}.
\end{split} 
\end{equation}

Similarly, the deflection in the vertical direction, $\delta_Y$ at point N is given by,
\begin{equation}   \label{Eq:delta_y}
    \begin{split} 
        \delta_Y &= \diffp[]{SE}W[W=0]\\
        &=\frac {3 (-b + \beta + l_2 \sin (\theta_2)) (a F + \alpha  + 
        Fl_2\cos (\theta_2) + \gamma F_S} {6YI (F\cos (\theta_2) - 
      F_S\cos (\theta_3 - \theta_2))}\\
        &\frac{- l_2 F_S\cos (\theta_3 - \theta_2))^2 - 
    3 (\beta - b) (a F + \alpha F + \gamma F_S)^2} {6YI (F\cos (\theta_2) - 
      F_S\cos (\theta_3 - \theta_2))}\\
      &- \frac {\sin (\theta_2) (a F + \alpha F + 
         Fl_2\cos (\theta_2) + \gamma F_S - l_2 F_S\cos (\theta_3 - \theta_2))^3}{6YI (F\cos (\theta_2) - 
       F_S\cos (\theta_3 - \theta_2))^2} \\
       &-\frac{\sin (\theta_2) (a F + \alpha F + \gamma F_S)^3)} {6YI (F\cos (\theta_2) - 
       F_S\cos (\theta_3 - \theta_2))^2} \\
       &+ \frac {3 (\beta + l_1\cos (\theta_1)) (\alpha F + 
        Fl_1\sin (\theta_1) + \gamma F_S - l_1 F_S\sin (\theta_1 + \theta_3))^2}{6YI (F\sin (\theta_1) - 
      F_S\sin (\theta_1 + \theta_3))} \\
      &- \frac{3\beta (\alpha F + \gamma F_S)^2} {6YI (F\sin (\theta_1) - 
      F_S\sin (\theta_1 + \theta_3))} \\
      &- \frac {\cos (\theta_1) ((\alpha F + 
         Fl_1\sin (\theta_1) + \gamma F_S - l_1 S\sin (\theta_1 + \theta_3))^3}{6YI (F\sin (\theta_1) - 
       F_S\sin (\theta_1 + \theta_3))^2}\\
       &-\frac{ \cos (\theta_1)(\alpha F + \gamma F_S)^3 } {6YI (F\sin (\theta_1) - 
       F_S\sin (\theta_1 + \theta_3))^2}.
    \end{split}
\end{equation}
\subsection{Stretch Ratio}
We define the stretch ratio, $\mu$, as the ratio of horizontal displacement to the vertical displacement,
\begin{equation}
    \mu = \frac{\delta_X}{\delta_Y}.
    \label{Eq:mudef}
\end{equation}
In the subsequent sections, we refer to stretch ratio while quantifying the Poisson's ratio mimicked by the mechanism. Note that for the positive Poisson's ratio effect, $\mu$ of the mechanism would be negative and vice versa. 
\subsubsection{\texorpdfstring{Positive $\mu$}{Positive mu}}
For smaller values of $K$, $\mu$ will be positive ($\mu_p$) whereas for larger values of $K$, it will be negative ($\mu_n$).When stiffness is zero ($K=0$),  $\mu$ is positive. To obtain an expression for $\mu_p$,  we set $F_S=0$  in Eqs. \ref{Eq:delta_y} and \ref{Eq:delta_x}, we have the horizontal displacement,
\begin{equation}
    \begin{split}
 \delta_{xp}&=\frac{\csc (\theta_1) \left(3 (\alpha+l_1 \sin (\theta_1)) (\alpha F+F l_1 \sin (\theta_1))^2-3 \alpha^3 F^2\right)}{6FYI}\\
 &-\frac{\csc (\theta_1) \left((\alpha F+F l_1 \sin (\theta_1))^3-\alpha^3 F^3\right)}{6F^2 YI}\\
 &-\frac{\sec (\theta_2) ((\alpha F+aF +F l_2 \cos (\theta_2))^3}{6F^2YI}\\
 &-\frac{\sec (\theta_2)(\alpha F+a F)^3}{6F^2YI}\\
 &+\frac{3\sec (\theta_2)(\alpha)(\alpha F+a F+F l_2 \cos (\theta_2))^2}{6FYI}\\
 &+\frac{3\sec (\theta_2)(a)(\alpha F+aF+F l_2 \cos (\theta_2))^2}{6FYI}\\  &+\frac{3\sec (\theta_2)l_2 \cos (\theta_2)(\alpha F+aF+F l_2 \cos (\theta_2))^2}{6FYI}\\
 &-\frac{3\sec (\theta_2) (\alpha+a) (\alpha F+aF)^2}{6FYI},
    \end{split}
\end{equation}
and vertical displacement,
\begin{equation}
    \begin{split}
\delta_{yp}&=-\frac{3 (\beta-b) (\alpha F+a F)^2)}{6FYI}\\
&-\frac{\cot (\theta_1) \csc (\theta_1) \left((\alpha F+F l_1 \sin (\theta_1))^3-\alpha^3 F^3\right)}{6F^2YI}\\
&+\frac{3\csc (\theta_1) \left(\beta+l_1 \cos (\theta_1) (\alpha F+F l_1 \sin (\theta_1))^2-3 \alpha^2 \beta F^2\right)}{6FYI}\\
&+\frac{3\sec (\theta_2)(-b+\beta)(\alpha F+a F+Fl_2 \cos (\theta_2))^2}{6FYI}\\
&+\frac{3\sec (\theta_2)(l_2 \sin (\theta_2)) (\alpha F+aF +Fl_2 \cos (\theta_2))^2}{6FYI}\\
&-\frac{\tan (\theta_2) \sec (\theta_2) ((\alpha F+aF +F l_2 \cos (\theta_2))^3}{6F^2YI}\\
&-\frac{\tan (\theta_2) \sec (\theta_2)(\alpha F+a F)^3}{6F^2YI}.
    \end{split}
\end{equation}
Thus, the positive stretch ratio is,
\begin{equation}
    \mu_p = \frac{\delta_{xp}}{\delta_{yp}}.
    \label{Eq:pSR}
\end{equation}

\subsubsection{\texorpdfstring{Negative $\mu$}{Negative mu}}

\begin{figure}[htbp]
    \centering
    \includegraphics[width=0.35\linewidth]{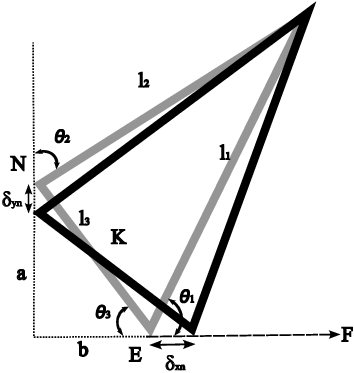}
    \caption{Quarter of the mechanism with negative $\mu$.}
    \label{fig:negative_SR}
\end{figure}

We observed in the previous section that when the spring force is zero, $\mu$ is positive. On the other hand, when the stiffness of the spring is large, $\mu$ would be negative. To obtain the deformations in this case, we can apply the principle of superposition by considering sub-problems with and without spring force, and combining them with a compatibility equation. However by  considering that $K$ is large we assume a rigid connection between points N and E as shown in Fig. \ref{fig:negative_SR}.
In this case, the displacement of point E outwards will cause point N to move inwards due to the pulling effect of the rigid segment, which creates a positive Poisson's ratio effect.  We define the negative stretch ratio associated with this configuration as, $\mu_n$. 

By observing that the rigid segment is in-extensible, Fig. \ref{fig:negative_SR}, we have 
\begin{equation}
\delta_{xn}^2+\delta_{yn}^2 + 2a\delta_{yn} + 2b\delta_{xn}= 0   .
    \label{Eq:nSR_0}
\end{equation}
As long as the deformations at N and E are much smaller than the lengths $a$ and $b$, we can neglect the square terms in Eq. \ref{Eq:nSR_0}. That is, 
\begin{equation}
    a\delta_{yn}= -b\delta_{xn}\implies      \mu_n = \frac{\delta_{xn}}{\delta_{yn}} =-\frac{a}{b}  .
    \label{Eq:nSR}
\end{equation}

\section{Design and Validation }
\label{sec:validation}
The design problem is to the obtain the geometric parameters of the mechanism while $\mu_p$ and $\mu_n$ are prescribed. The geometric parameters involve, $a$, $b$, $\theta_1$, and $\theta_2$. Note that $l_1$ and $l_2$ are not independent variables, since 
\begin{align}
l_1&=\cosec (\theta_1) \left(\frac{\cos (\theta_2) (a \cot (\theta_1)+b)}{\sin (\theta_2)-\cot (\theta_1) \cos (\theta_2)}+a\right),\\
l_2&=\frac{a \cot (\theta_1)+b}{\sin (\theta_2)-\cot (\theta_1) \cos (\theta_2)}.
\label{Eq:l1l2}
\end{align}
When $\mu_n$ is specified, we see from Eq. \ref{Eq:nSR} that the ratio between $a$ and $b$ is fixed. Additionally, Eq. \ref{Eq:pSR} needs to satisfied for getting a particular $\mu_p$. Since there are four design variables and only two equations, multiple solutions are possible. 

\subsection{\texorpdfstring{Example I: $\mu_n=-4$, $\mu_p=2.1$}{Example I: mun=-4, mup=2.1}}

\begin{figure}[!htbp]
    \centering
    \includegraphics[width=0.5\linewidth]{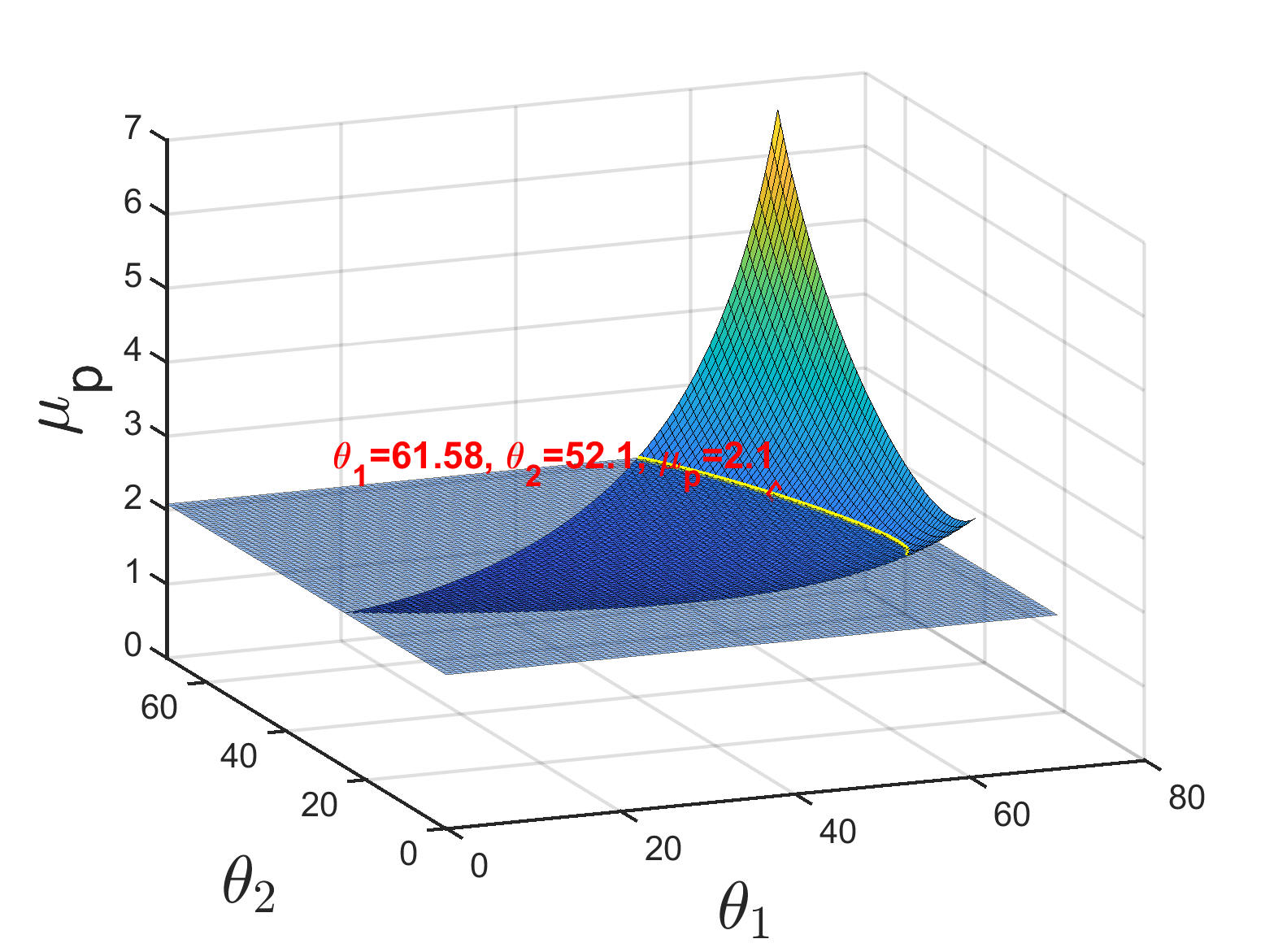}
    \caption{$\mu_p$ for varying $\theta_1$ and $\theta_2$ ($\mu_n=-4$)}
    \label{fig:SR4_surf}
\end{figure}

\begin{figure}[!htbp]
    \centering
    \includegraphics[width=0.5\linewidth]{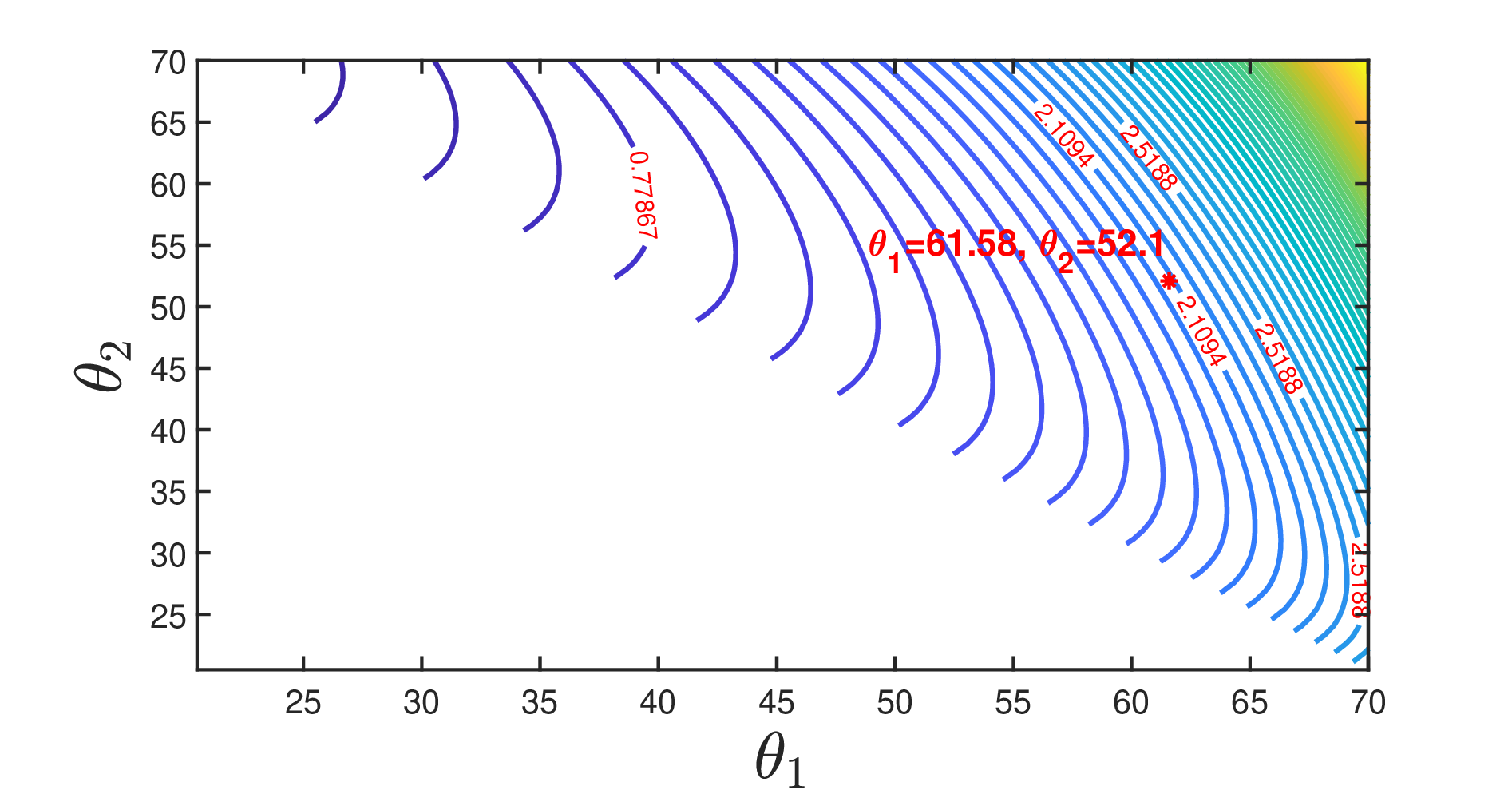}
    \caption{$\mu_p$ for varying $\theta_1$ and $\theta_2$ ($\mu_n=-4$)}
    \label{fig:SR4_contour}
\end{figure}

We start by deciding $a$ and $b$ so that the negative stretch ratio, $\mu_n$ is $-4$. We take $a$ as 25mm and $b$ as 6.25mm. By substituting these values in Eq. \ref{Eq:pSR} we get a design surface as shown in Fig. \ref{fig:SR4_surf}.
Each point on this surface represents a different $\mu_p$ and corresponding $\theta_1$ and $\theta_2$, while $\mu_n=-4$. By selecting a point on this surface, we can design a mechanism with any $\mu_p$ from 0.5 to 7, as indicated by the z-axis of Fig. \ref{fig:SR4_surf}. Figure \ref{fig:SR4_contour} shows a contour plot of the design surface.
Each contour represents a particular $\mu_p$ value, indicating multiple solutions. 
The range of  $\theta_1$ and $\theta_2$ is between a lower limit of $0^{\circ}$ and an upper limit of $75^{\circ}$. The upper and lower limits are set by considering the need to preserve the re-entrant shape of the mechanism. If both the angles are $90^{\circ}$ the shape of the mechanism will be rectangular. Hence we need to keep angles below  $90^{\circ}$. Furthermore, the sensitivity of $\mu_p$ is high when  $\theta_1$ and $\theta_2$ are close to $90^\circ$ as indicated by dense contour lines in Fig. \ref{fig:SR4_contour}. Hence, the limit of $75^{\circ}$ is chosen here with a factor of safety of the upper limit, $90^{\circ}$. It may be noted that there are no contours in the lower left triangular region of Fig. \ref{fig:SR4_contour}. This is to ensure that $\theta_1>90^\circ-\theta_2$, which is the condition  for the segments of $l_1$ and $l_2$ to meet in the first quadrant.  

By intersecting the design surface with a plane, $\mu_p=2.1$, we obtain a curve as shown in Fig. \ref{fig:SR4_surf}. To complete our design, we  take a point arbitrarily on this curve as indicated by the marker. The $\theta_1$ and $\theta_2$ of this point are 61.58$^{\circ}$and 52.1$^{\circ}$. We calculate corresponding $l_1$ and $l_2$ from Eq. \ref{Eq:l1l2}.

\subsection{\texorpdfstring{Example II: $\mu_n=-1$, $\mu_p=4.1$}
{Example II: mun=-1, mup=4.1}}

\begin{figure}[htbp!]
    \centering
    \includegraphics[width=0.5\linewidth]{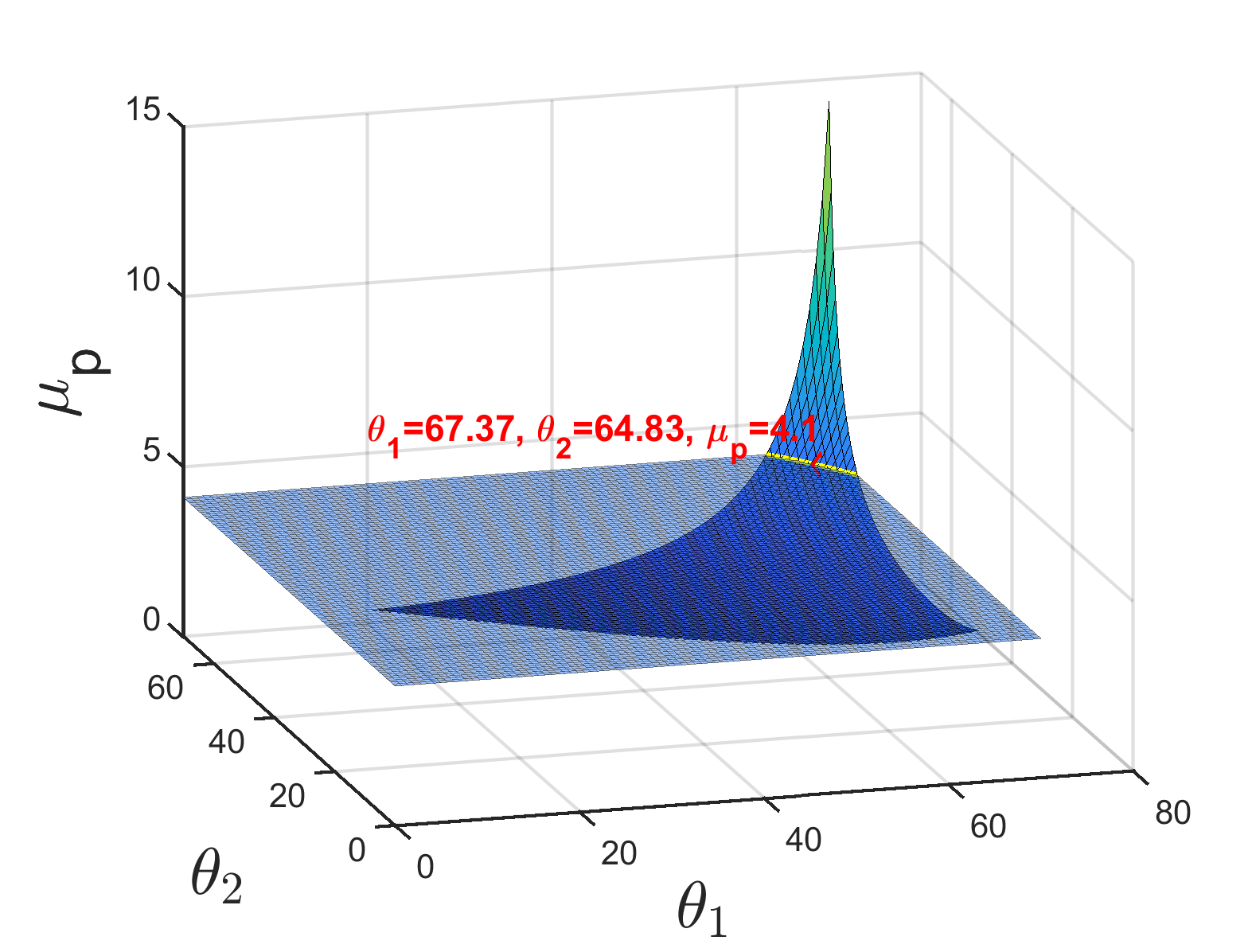}
    \caption{ $\mu_p$ for varying $\theta_1$ and $\theta_2$ ($\mu_n=-1$)}
    \label{fig:SR1_design}
\end{figure}

In Example II, let us consider a different set of stretch ratios---$\mu_n=-1$, $\mu_p=4.1$. The methodology to obtain the dimensions remains the same. A design surface is generated for $\mu_n=-1$, by varying the geometrical parameters, $\theta_1$ and $\theta_2$ as shown in Fig. \ref{fig:SR1_design}. 
The parameters $a$ and $b$ are taken to be 2.5 mm each.
By intersecting the design surface with a plane, $\mu_p=4.1$ and arbitrarily choosing a point on the intersecting curve, we get $\theta_1$, $\theta_2$, $l_1$, and $l_2$ as 67.37$^{\circ}$, 64.83$^{\circ}$, 4.95 mm, 4.87mm, respectively.
\subsection{Validation}
\begin{figure}[!htbp]
    \centering\includegraphics[width=0.5\linewidth]{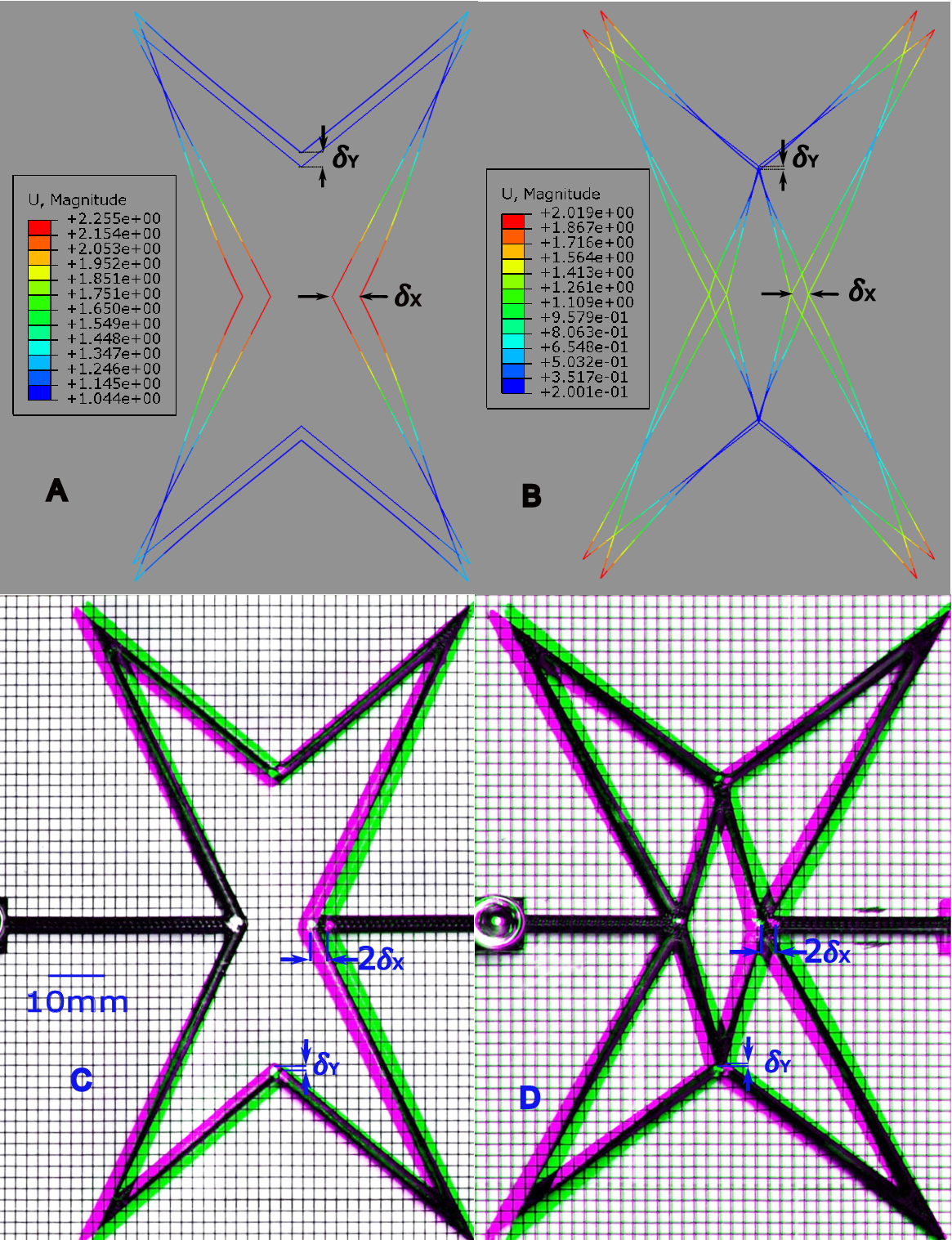}
    \caption{Deformed mechanisms superimposed over initial state from FEA (A and B) and experiments (C and D) for Example I. }
    \label{fig:validation}
\end{figure}

Finally, we validate the two designs obtained from  modelling and compare them to the results from FEA and experiments conducted on prototypes. As mentioned before,  two configurations of the mechanism are considered  separately. The design corresponding to $\mu_p$, where the $K=0$ is shown on the left in Fig. \ref{fig:validation}. 
The one corresponding to $\mu_n$, with a rigid segment in-between, is shown on the right. The  mechanism shown in Fig. \ref{fig:validation} corresponds to Example I.
\subsubsection{FEA}

The stretch ratios of the mechanisms are obtained from FEA and compared with  Eqs. \ref{Eq:nSR} and \ref{Eq:pSR}. We conducted a quasi-static study with a linear (NLGEOM-OFF) wire model of the frame with beam elements in Abaqus 2018. The cross-section of the mechanism is a square of 3mm side. The material used here is Onyx 3D printing material, which we later used for experimental verification. It has Young’s modulus, Y of 240 MPa, density of 1200 kg/m$^3$, and Poisson's ratio of 0.3. 

We give equal and opposite displacements at the two input points (E and W in Fig. \ref{fig:frame_with_springs}). Corresponding vertical displacement of the output point (N in Fig. \ref{fig:frame_with_springs}) is extracted. These values for Example I are shown in Figs. \ref{fig:validation} A and \ref{fig:validation} B. They are also summarized in Table \ref{tab:comparison} for both examples.
\begin{table}[!htbp]
\caption{Comparing analytical model with FEA and experimental results}
\label{tab:comparison}
\centering
\small
\setlength{\tabcolsep}{3pt}
\begin{tabular}{ccccccccccc}
\toprule
\multicolumn{2}{c}{\multirow{2}{*}{Design}} &
\multirow{2}{*}{\begin{tabular}[c]{@{}c@{}}Analytical\\ SR\end{tabular}} &
\multirow{2}{*}{\begin{tabular}[c]{@{}c@{}}X-EXP\\ (Input)\end{tabular}} &
\multirow{2}{*}{Y-EXP} &
\multirow{2}{*}{SR-EXP} &
\multirow{2}{*}{\begin{tabular}[c]{@{}c@{}}X-FEM\\ (Input)\end{tabular}} &
\multirow{2}{*}{Y-FEM} &
\multirow{2}{*}{SR-FEM} &
\multirow{2}{*}{\begin{tabular}[c]{@{}c@{}}\% Error\\ FEA-Analytical\end{tabular}} &
\multirow{2}{*}{\begin{tabular}[c]{@{}c@{}}\% Error\\ Exp-Analytical\end{tabular}} \\
\multicolumn{2}{c}{} & & & & & & & & & \\
\midrule
\multirow{2}{*}{1} & $\mu_p$ & 2.1 & 2.255 & 1.064 & 2.119 & 2.255 & 1.082 & 2.084 & 0.762 & 0.905 \\
                   & $\mu_n$ & -4  & 1.355 & -0.330 & 4.106 & 1.355 & -0.329 & 4.119 & 2.994 & 2.651 \\
\multirow{2}{*}{2} & $\mu_p$ & 4.1 & 0.883 & 0.210 & 4.205 & 0.883 & 0.216 & 4.084 & 0.527 & 2.405 \\
                   & $\mu_n$ & -1  & 0.405 & -0.396 & 1.022 & 0.405 & -0.397 & 1.019 & 1.954 & 2.272 \\
\bottomrule
\end{tabular}
\end{table}
The errors are within $3\%$ for both the examples. The comparitively larger errors observed in $\mu_n$ is due to the approximation in Eq. \ref{Eq:nSR}.
\subsubsection{Experiments}
The mechanisms are 3D-printed in Mark 2, Markforged printer using Onyx material. The Onyx Material was chosen because of its distortion-free and anti-warping properties when 3D printed, and it is flexible enough to apply displacement without much effort.

For the experiment, one of the input points (W in Fig. \ref{fig:frame_with_springs}) was kept fixed, and a horizontal displacement was given at the other end, as shown in Fig. \ref{fig:validation}. This is done to ease applying the input load while conducting the experiments and should not alter the deformation in the transverse direction. The horizontal load was applied using a linear stage to the extensions on the model. The vertical displacement was measured at the output, and the ratio of half the horizontal displacement to the corresponding vertical displacement was calculated to find the $\mu_p$ and $\mu_n$. 

The calculations are based on image processing with the aid of the  ImageJ software. Figure \ref{fig:validation} shows the superimposed images of the mechanism before(pink) and after(green) the displacement has been applied. The $\mu_p$ and $\mu_n$ from experimental observations are tabulated and compared with the analytical model in Table \ref{tab:comparison}. We observe that $\mu_p$ and $\mu_n$ from modelling, FEA, and experiments agree.

\section{Summary and future work}
\label{sec:summary}
We presented the design of tunable compliant micro-mechanisms with invertible and tunable Poisson's ratio effect. We intend to use these mechanisms to study the behaviour of biological cells dependent on the mechanical properties of the substrates they are growing on. 

The mechanical design is based on incorporating a stiffness-variable spring in between a re-entrant structure. By flipping the stiffness of the springs to their extremes, the design can have negative and positive stretch ratios with the desired magnitude. These two cases are analytically modelled here by assuming the bistable element as a spring of infinite and zero stiffness, respectively. The utility of the modelling was illustrated by examples and verified using FEA and tabletop experiments. 

{At the micro-scale, the dimensions of the mechanism will be in the order of microns---for example, 5 $\mu$m $\times$ 5 $\mu$m cross-section and tens of $\mu$m long. The mechanism will be microfabricated out of SU-8 using a two-layer lithography process we have previously reported for fabricating a uniaxial single-cell stretcher} \cite{Kollimada2017}. {At these dimensions, the kinematic relationships of the macro- and micro-sized mechanism are comparable since Euler-Bernoulli beam theory is valid at both these scales.}

\begin{figure}[!htbp]
    \centering
    \includegraphics[width=0.45\linewidth]{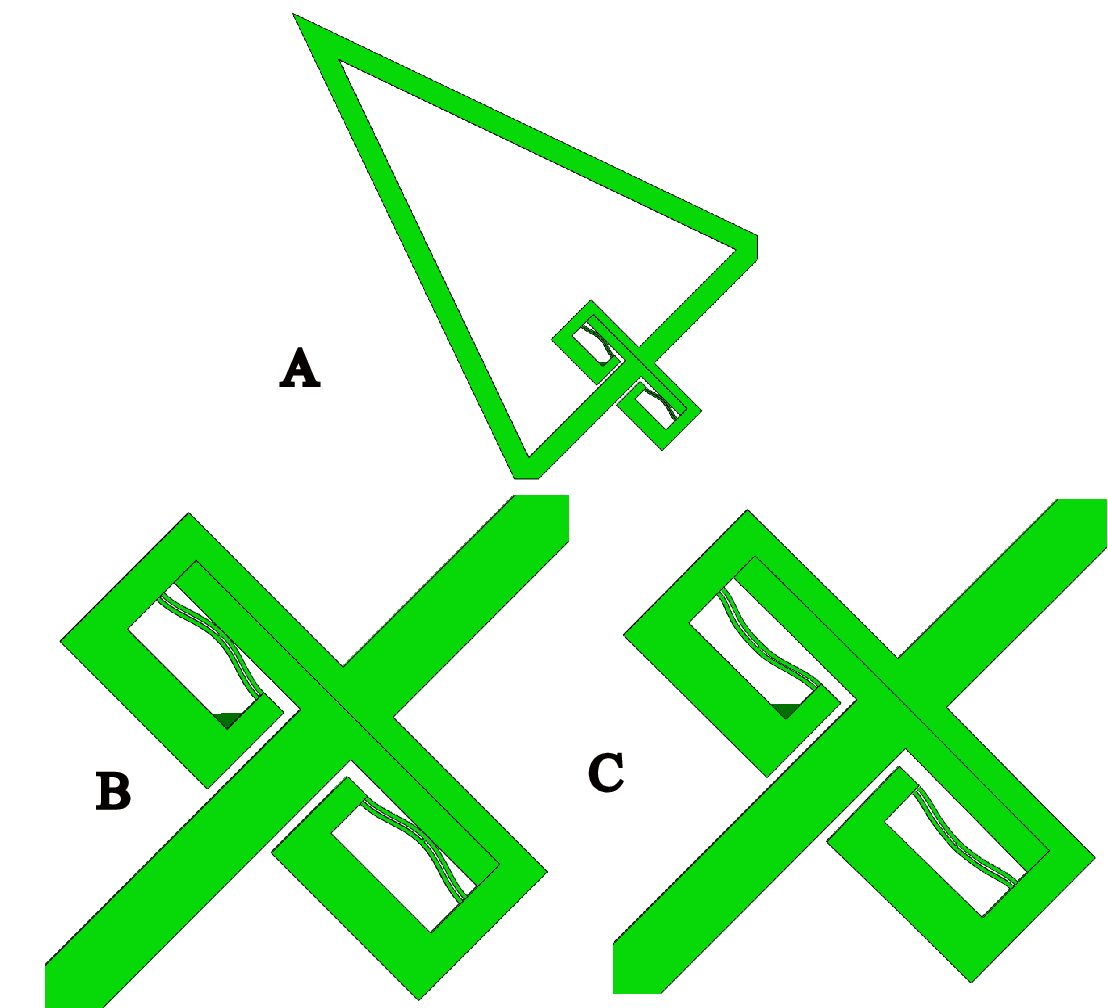}
    \caption{ (A) Quarter of the mechanism with a bistable element. (B) Bistable mechanism engaged ($\mu_n$) (C) Bistable mechanism disengaged ($\mu_p$)}
    \label{fig:bistable_element}
\end{figure}

A bistable element could be used as a tunable spring element. A quarter of the mechanism with a preliminary design of the bistable element is shown in Fig. \ref{fig:bistable_element} A.
The configuration in Fig. \ref{fig:bistable_element} B corresponds to the engaged state giving a negative stretch ratio. The one in Fig. \ref{fig:bistable_element} C is the disengaged state with a positive stretch ratio. The design of bistable arches is well studied in the literature\cite{palathingalJMR,PalathingalIJSS,PalathingalMMT}. We will incorporate bistable arches into our design as the next step in this work. Furthermore, we will microfabricate the mechanism, grow cells on it and study the effect of inverting the Poisson's ratio on cell function.


\section*{Acknowledgments}
We gratefully acknowledge the funding received from the Department of Science and Technology, Government of India (SRG/2021/001463) and IIT-H SG-92.

\nocite{*}
\bibliographystyle{plain}
\bibliography{references}

\vskip2pc

\end{document}